\documentclass[preprint]{iucr}              

     \journalcode{J}              
\newcommand{\bm}[1]{{\mbox{\boldmath $#1$}}}
\usepackage{amsmath}
\usepackage{breqn}
\usepackage{graphicx}
\usepackage[dvipdfmx]{color}
\usepackage{comment}
\usepackage{color} 

\begin{document}                  



\title{Orientation mapping of YbSn$_3$ single crystals based on Bragg-dip analysis using a delay-line superconducting sensor}

\cauthor[a, b]{Hiroaki}{Shishido}{shishido@omu.ac.jp}{}
\author[c, d]{The Dang}{Vu}
\author[d]{Kazuya}{Aizawa}
\author[c, e]{Kenji M.}{Kojima}
\author[c]{Tomio}{Koyama}
\author[d]{Kenichi}{Oikawa}
\author[d]{Masahide}{Harada}
\author[d]{Takayuki}{Oku}
\author[d]{Kazuhiko}{Soyama}
\author[f]{Shigeyuki}{Miyajima}
\author[g]{Mutsuo}{Hidaka}
\author[h]{Soh Y.}{Suzuki}
\author[i]{Manobu M.}{Tanaka}
\author[c]{Shuichi}{Kawamata}
\author[c]{Takekazu}{Ishida}

\aff[a]{Department of Physics and Electronics, Graduate School of Engineering,Osaka Metropolitan University, Sakai, Osaka 599-8531, Japan}
\aff[b]{The Center for Research and Innovation in Electronic Functional Materials, Osaka Metropolitan University, Sakai, Osaka 599-8570, Japan}
\aff[c]{Division of Quantum and Radiation Engineering, Osaka Metropolitan University, Sakai, Osaka 599-8570, Japan}
\aff[d]{Materials and Life Science Division, J-PARC Center, Japan Atomic Energy Agency, Tokai, Ibaraki 319-1195, Japan}
\aff[e]{Centre for Molecular and Materials Science, TRIUMF and Stewart Blusson Quantum Matter Institute, University of British Columbia, Vancouver, BC, V6T 2A3 and V6T 1Z4, Canada}
\aff[f]{Advanced ICT Research Institute, National Institute of Information and Communications Technology, 588-2 Iwaoka, Nishi-ku, Kobe, Hyogo 651-2492, Japan}
\aff[g]{Advanced Industrial Science and Technology, Tsukuba, Ibaraki 305-8568, Japan}
\aff[h]{Computing Research Center, Applied Research Laboratory, High Energy Accelerator Research Organization (KEK), Tsukuba, Ibaraki 305-0801, Japan}
\aff[i]{Institute of Particle and Nuclear Studies, High Energy Accelerator Research Organization (KEK), Tsukuba, Ibaraki 305-0801, Japan}








\maketitle                        


\begin{abstract}
Recent progress in high-power pulsed neutron sources has stimulated the development of the Bragg-dip and Bragg-edge analysis methods using a two-dimensional neutron detector with high temporal resolution to resolve the neutron energy by the time-of-flight method.
The delay-line current-biased kinetic-inductance detector (CB-KID) is a two-dimensional superconducting sensor with a high temporal resolution and multi-hit capability.
We demonstrate that the delay-line CB-KID with a $^{10}$B neutron conversion layer can be applied to high-spatial-resolution neutron transmission imaging and spectroscopy up to 100\,eV.
Dip structures in the transmission spectrum induced by Bragg diffraction and nuclear resonance absorption in YbSn$_3$ single crystals. 
We successfully drew the orientation mapping of YbSn$_3$ crystals based on the analysis of observed Bragg-dip positions in the transmission spectrum.
\end{abstract}


\section{Introduction}

The transmission capability of neutron beams is highly different from that of X-rays; thus, neutron imaging can be crucial in utilizing non-destructive inspection techniques \cite{Leh15}. 
Neutron imaging has a notable advantage in the high penetration of neutron beams and its ability to observe rich information on material structures through the variation of the transmission coefficient as a function of neutron energy.

The dip structure induced by the nuclear resonance absorption appearing in the transmission spectrum allows us to perform elemental discriminating imaging, revealed as neutron-resonance transmission imaging \cite{Kai13, Tre14, Fes15, Tre17AIP, Tre17SR} and tomography \cite{Vog20, Los22}.
One can also obtain information on crystal structure, crystallinity, crystal grain, lattice strain, and grain orientation from the dip structures induced by the Bragg diffraction.   
Polycrystalline samples exhibit saw-tooth-like dip structures called Bragg edges in the transmission spectrum \cite{Fer47}.  
Edge positions as a function of wavelength are determined by the distance between lattice planes. 
The shift in edge position, edge broadening, and overall edge shape reflect strains, preferred orientation, and crystallite size, respectively  \cite{San02, Sat11, Sat15}.

Sharp dips, called the Bragg dips, appear in the transmission spectrum of single crystalline samples \cite{Mal16, Tre17SR}. 
In sharp contrast with polycrystalline samples, the Bragg diffraction occurs at the specific wavelength determined by the Miller indices and single crystal orientation.  
One can determine the lattice constant, crystal orientation, and mosaicity of the crystals by a profile fitting of a Bragg-dip pattern \cite{San05, Mal16 }. 
Additionally, a mapping of the crystallographic direction of grains with respect to the incident beams is possible using Bragg-dip analyses, facilitated by a nuclear-spallation pulsed neutron source \cite{Sat17}.
The reconstruction of three-dimensional mapping of the crystallographic direction of grains with a spatial resolution of several hundred micrometers was also examined based on the neutron diffraction \cite{Los22}.
Bragg-dips' and edges' depths decrease with higher order Miller indices or higher energy region, and are therefore usually visible in a limited neutron energy range.  On the other hand, most of resonance-dip appear at neutron energies higher than $\sim$0.5\,eV.

Detectors with spatial and temporal resolutions have been intensively developed to apply neutron energy-resolved high-resolution imaging and Bragg-edge/-dip-based imaging analysis, such as micro-channel-plate detector \cite{Tre11}, Li-glass scintillator pixel-type detector \cite{Sat10}, gas electron multiplier \cite{Uno12}, and micro-pixel chamber-based neutron imaging detector \cite{Par16}.

We have developed a superconducting two-dimensional neutron detector with high spatial and temporal resolutions by combining a current-biased kinetic-inductance-detector (CB-KID) and a $^{10}$B neutron conversion layer \cite{Ish14, shi15}. 
The CB-KID operation is based on a transient change of the Cooper pair density induced by the nuclear reaction between an incident neutron and a $^{10}$B nucleus in the neutron conversion layer. 
It works over a wide temperature range below the superconducting transition temperature of Nb at high operating speed and has a spatial resolution of 16\,$\mu$m through a delay-line technique (delay-line CB-KID) \cite{Iiz19}.

The delay-line CB-KID also has multi-hit capability. 
It has two orthogonally stacked Nb meander lines. 
A neutron irradiation event simultaneously induces voltage signals on both meander lines. 
Signals propagate through meander lines as electromagnetic waves, and are finally detected by a time-to-digital converter Kalliope-DC \cite{Koj14, shi18} with a temporal resolution of 1\,ns.  
An event occurrence time $t_0$ can be estimated independently from signal arrival times of both meander lines. Therefore, multiple simultaneous events can be discriminated by a difference in $t_0$. 
If the $t_0$ difference is within 10\,ns, we regard that signals are induced by the same event. 
The signal discrimination limit, which determines the maximum multi-hit rate, is simply estimated to be about ten megahertz because the typical full width of half maximum of detected signals is 40\,ns.

In this study, we demonstrate that the delay-line CB-KID can be applied to the Bragg-dip-based imaging and analysis of a test sample consisting of a few single crystals. 
We succeeded in completing orientation mapping for YbSn$_3$ single crystals. 
We also show that nuclear resonance absorption analysis is possible in the energy range from 0.1 to 100\,eV with the CB-KID, by observing dips in the transmission spectrum that are consistent with nuclear resonance absorption and the natural abundance of Yb nuclei in YbSn$_3$, and can be used to calibrate the origin of a constant delay time $t_{\rm delay}$ for the time of flight (ToF).  
We note that 100\,eV is not high enough for most nuclei, while it is enough to identify resonance-dips of Yb nuclei.

\section{Experimental methods}

\subsection{Neutron imaging with pulsed neutrons}
\begin{figure}
\begin{center}
\includegraphics[width=0.8\linewidth, pagebox=cropbox, clip]{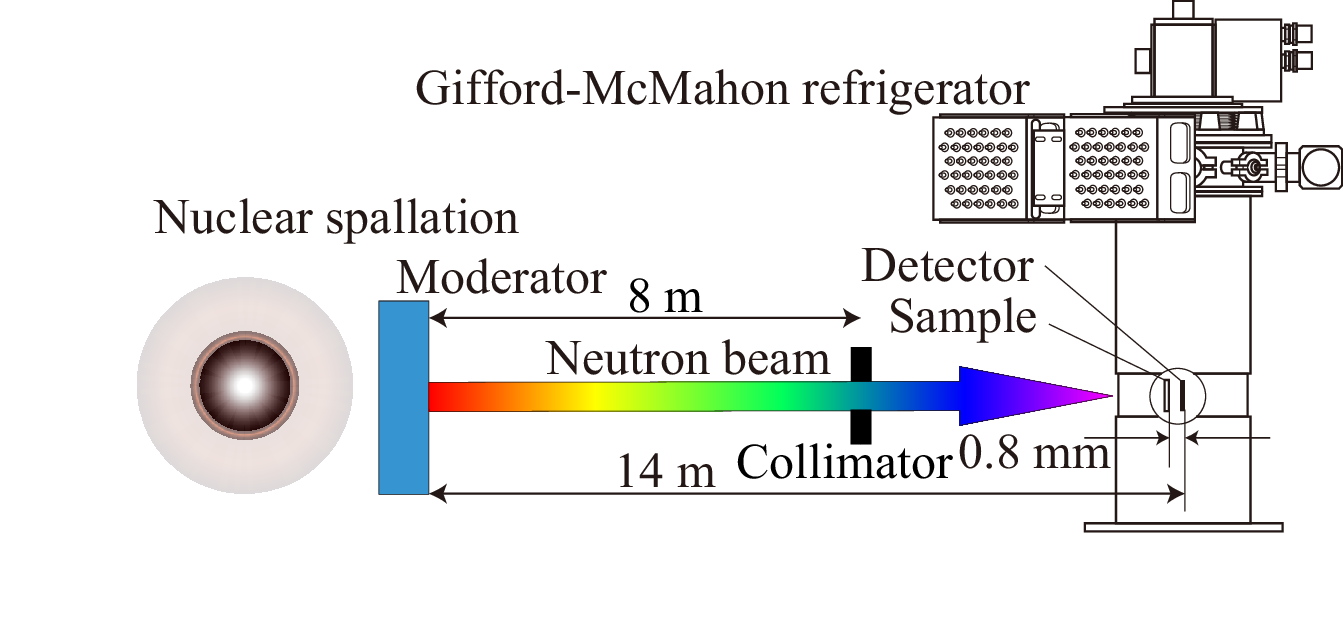}
\end{center}
\caption{Schematic illustration of the measurements system in the beamline at J-PARC (BL10). Pulsed neutrons traveled from the source moderator through the beamline toward the detector through an YbSn$_3$ sample. The collimator was installed 8\,m downstream from the moderator. The detector and sample were cooled down to 7.9\,K by using a Gifford-McMahon refrigerator.
}
\label{BeamLine}
\end{figure}

We performed neutron imaging experiments with pulsed neutrons at BL10 of the Material and Life science experimental Facility (MLF), J-PARC \cite{BL10}. 
Neutrons of various energies are created by the nuclear spallation reaction and travel from a moderator to the detector through the beamline with the length of 14\,m at BL10, as shown in Fig.~\ref{BeamLine}. 
Nuclear spallation reactions are initiated every 40\,ms and synchronized trigger signals are fed from the facility with a possible constant delay time $t_{\rm delay}$. 
This delay time is not negligible in treating neutrons with higher energies or shorter wavelengths.
Thus, we obtain a neutron wavelength $\lambda$ from ToF ($t$) by 
\begin{equation*}
\lambda = \frac{2\pi\hbar}{m_{\rm n} L}(t-t_{\rm delay}), 
\label{lambda}
\end{equation*}
where $m_{\rm n}$ is a neutron mass and $\hbar$ is the reduced Plank constant. 
The neutron energy $E$ is also represented as

\begin{equation*}
E=\frac{m_{\rm n}}{2}\left(\frac{L}{t-t_{\rm delay}}\right)^2.   
\label{t_delay}
\end{equation*}

We conducted imaging experiments by selecting the collimation ratio $L/D$=337 at a beam power of 920\,kW, where $D$ is the collimator aperture side length of 17.8\,mm, $L$ is the distance between the aperture and the detector of 6\,m. 
This is just under the maximum accelerator power of the J-PARC facility.

Neutron beam was detected by a delay-line CB-KID, which is a $^3$He-free solid-state neutron detector.   The detector was cooled down to  $T=7.9$\,K to keep it in a superconducting state by using a Gifford-McMahon refrigerator \cite{McM60}, as shown in Fig.~\ref{BeamLine}.  
During the measurements, DC bias currents of 30\,$\mu$A were fed to the detector.   
Details of the detector are described in our previous work \cite{shi18}.

The test-sample temperature can be kept at room temperature, although the distance between the test sample and the detector would then be more than 2\,cm.   Generally, longer distance between the sample and detector degrades the spatial resolution of the transmission image.    
For example, the spatial resolution is reduced to 60\,$\mu$m or more at $L/D$=337. 
In this experiments, the test sample was also cooled down to 7.9\,K to place at a distance of 0.8\,mm in front of the detector for achieving higher spatial resolution. 

\subsection{Preparation for YbSn$_3$ samples}

We used a metallic compound YbSn$_3$ as a test sample. 
YbSn$_3$ is known as a divalent Yb compound with pocket Fermi surfaces \cite{Sch77, Saka97}.

It crystallizes in the AuCu$_3$-type simple cubic structure (space group $Pm\bar{3}m$ \#221) with the lattice constant $a$ = 0.468\,nm \cite{Pal75}. 
We grew YbSn$_3$ single crystals using the Sn-self-flux method. 
The starting materials of Yb and Sn with an atomic ratio of 1:4  were sealed in a quartz tube under Ar atmosphere.
This tube was heated up to 800\,$^\circ$C and cooled down to 220\,$^\circ$C with a cooling rate of 5\,$^\circ$C/h. 
A platelet sample with a thickness of 1.6\,mm was cut out from the obtained ingot.

By considering neutron-absorption channels in YbSn$_3$ crystals, we note that the natural abundance of Yb isotopes is 0.140\%, 3.04\%, 14.31\%, 21.82\%, 16.13\%, 31.84\%, and 12.73\%  for $^{168}$Yb, $^{170}$Yb, $^{171}$Yb, $^{172}$Yb, $^{173}$Yb, $^{174}$Yb, and $^{176}$Yb, respectively\cite{Wan15}.

\subsection{Procedure to assign Bragg-dip indices}
Assignment procedures of the Bragg dip was described in details in the preceding reports. 
Santisteban (2005) assigned Miller indexes of Bragg dips based on the crystallographic geometrical calculations.
Sato {\it et al.}, (2017) assigned them based on the matching analyses with a database of various Bragg-dip patterns.

\begin{figure}
\begin{center}
\includegraphics[width=0.5\linewidth, pagebox=cropbox, clip]{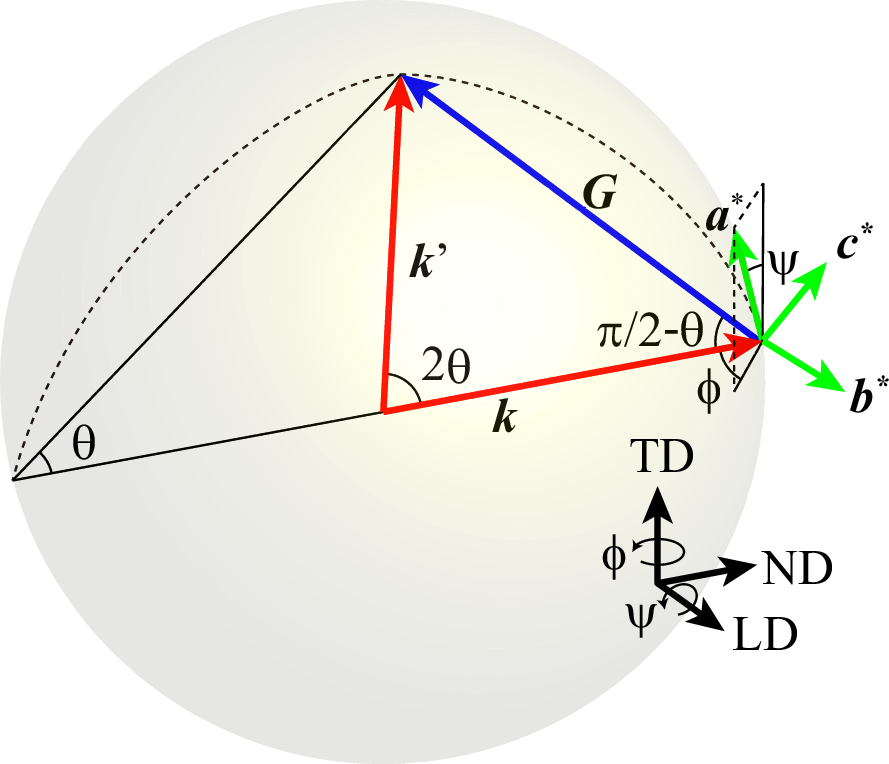}
\end{center}
\caption{A schematic view of the Ewald sphere with wavevectors of incoming beams {\bm k}, outgoing beams {\bm k}', reciprocal lattice vector {\bm G}, and reciprocal lattice unit vector {\bm a}$^*$. {\bm a}$^*$ is rotated by an azimuthal angle $\phi$ and a longitudinal angle $\psi$ from the vertical line, and is perpendicular to {\bm k} when $\psi=0$. 
In the figure, LD and TD are transverse directions of neutron beams, and ND is the normal direction (or the direction of neutron beams).}
\label{Ewald}
\end{figure}

We employed an assignment procedure based on the crystallographic geometrical calculation.  
The assignment of the Bragg dip was carried out by the following steps:
(1) A maximum wavelength of a Bragg-dip position is limited to $\lambda_{hkl} \leq 2d_{hkl}$, where $\lambda_{hkl}$ is a Bragg-dip position for $hkl$ and equivalent Bragg reflections and $d_{hkl}$ is interplanar distance for ($hkl$) and equivalent planes, where $h$, $k$, and $l$ are Miller indices \cite{San05}.  
In the simple cubic structure case,  $d_{100}$ and $d_{110}$ are the longest and second longest, respectively.   
If a Bragg dip appears at $\lambda > 2d_{110}$, one can assign it as the 100 Bragg dip because no Bragg dips appear above $\lambda > 2d_{110}$ except the Bragg dip corresponding to the 100 and equivalent diffraction dips.   
(2) Using similar considerations comparing with the third longest interplanar distance of $d_{111}$, the 110 Bragg dip is also assigned. 
(3) A crystal rotation angle $\psi$ about LD-direction and $\phi$ about TD-direction shown in Fig.~\ref{Ewald} are determined even with just $\lambda_{100}$ and $\lambda_{110}$ observed simultaneously with neutrons incident from one direction of the crystal. 
A crystal rotation angle about the neutron beams or normal direction (ND) can not be determine.   
(4) Calculate all Bragg dips based on $\psi$ and $\phi$ of the single crystal for comparison with the experimental results.
The assignment is deterministic if 100 and 110 Bragg dips appear at $\lambda_{100} > 2d_{110}$ and $\lambda_{110} > 2d_{111}$, respectively.   In other cases, the assignment may have several choices for assigning 100 and 110 Bragg dips.  In such cases, one should examine one of possible combinations, and compare calculated dips with experimental dips. 
If one fails to assign the Bragg-dip index, change the 100-dip assignment and/or the 110-dip assignment to recalculate the Bragg-dip assignments.

We calculated the $\psi$ and $\phi$ angles of the reciprocal lattice unit vector $\bm{a}^*$ from the 100 and 110 Bragg dips, as described below. 
Figure~\ref{Ewald} shows a schematic view of the Ewald sphere with a wavelength of $\lambda$ in the reciprocal space. 
According to the expression given by Santisteban (2005), $\lambda$ is given in the present case as follows:
\begin{equation}
		\lambda_{hkl}=\frac{2a(-h\cos\psi\cos\phi+k\sin\phi-l\sin\psi\cos\phi)}{h^2+k^2+l^2}.
		\label{eq_lambda}
\end{equation}
Then, we obtain the formula of $\lambda(1, 0, 0)$ and $\lambda(1, 1, 0)$ from eqn.~(\ref{eq_lambda}) as follows:
\begin{equation}
\left\{ \,
    \begin{aligned}
    & \lambda_{100} = -2a\cos\psi\cos\phi \\
    & \lambda_{110} = a(-\cos\psi\cos\phi+\sin\phi). 
	\label{eq_psi}
    \end{aligned}
\right.
\end{equation}
One can determine $\psi$ and $\phi$ by solving simultaneous eqns.~(\ref{eq_psi}) as follows:
\begin{align}
& \phi=\arcsin\left[\frac{\lambda_{110}}{a}-\frac{\lambda_{100}}{2a}\right], 
	\label{eq_phi+psi1} \\
& \psi=\arccos\left[-\frac{\lambda_{100}}{2a\cos{\phi}}\right].
	\label{eq_phi+psi2}
\end{align}
The wavelengths $\lambda_{hkl}$ for other indices can be calculated by substituting the determined $\psi$ and $\phi$ values (from eqn.~(\ref{eq_phi+psi1}) and eqn.~(\ref{eq_phi+psi2})) into eqn.~(\ref{eq_lambda}).

\section{Results and discussion}

\subsection{Neutron transmission image}

\begin{figure}
\begin{center}
\includegraphics[width=0.8\linewidth, pagebox=cropbox, clip]{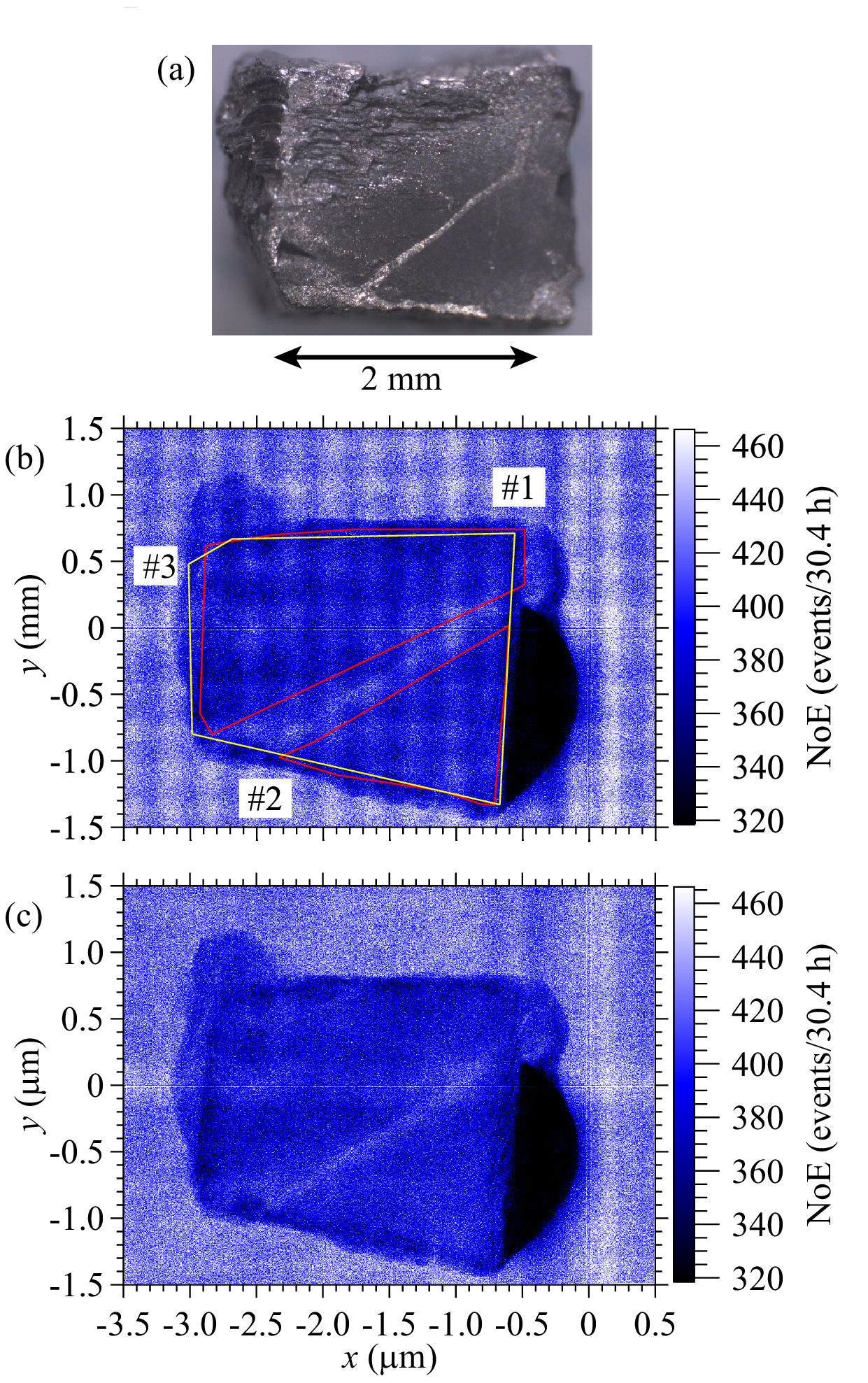}
\end{center}
\caption{(a) Optical photograph of the test sample. 
(b) Neutron transmission image of the YbSn$_3$ single crystals (\#1, \#2 in red) with the neutron energy ranging from 0.6 to 300\,meV.
The area \#3 in yellow is the region used in resonance absorption (see Fig.~\ref{Trans_logE}).  (c) A fast Fourier transformation filtered image. 
}
\label{Img}
\end{figure}

Figure~\ref{Img}(a) shows an optical photograph of the test sample. 
It consists of two YbSn$_3$ single crystals seen as dark-gray regions.
Two separate crystals are stuck together by some of the excess Sn. 
However, it is not very clear whether there are two crystals or not in the photograph, while a boundary can be seen as a bright diagonal line of Sn. 
Figure~\ref{Img}(b) shows the neutron transmission image with the color intensity scale for the number of events (NoE) taken over the neutron energy range of 0.6 $\leq E \leq$ 300\,meV accumulated for 30.4\,hours.  
The YbSn$_3$ single crystals are seen as dark-blue regions because the total cross section of Yb nuclei is higher than that of Sn. 
We note that the arc-shaped dark region on the right is an epoxy resin used to fix the sample on the holder in the minimum area. 
The trapezoidal YbSn$_3$ single crystal (\#1) on the upper left and the triangular one (\#2) on the lower right are stuck together by a thin layer of Sn.
The minimum width of Sn layer is less than 100\,$\mu$m, and may not be distinguished with $L/D$ = 337 if the sample is placed on the outside of the chamber.
Pixel sizes have incomplete periodicity.
Although most pixels of the image have the sizes of 6 $\times$ 4.5\,$\mu$m$^2$, a few pixels have the sizes of 4.5$\times$4.5, 6$\times$6, and 4.5$\times$6\,$\mu$m$^2$. 
These pixel sizes are smaller by an order of magnitude than those reported in the previous research \cite{Tre11}. 
A thin white grid pattern can be seen in the image, but it is an artifact resulting from nonuniformity in pixel sizes in the analysis. 
Because the total number of events for pixels with larger sizes is larger than that for pixels with smaller sizes.  In fact, grid lines appear nearly periodic, matching the rough periodicity of pixel size variation, while the grid pattern is blurred due to the limitation of the spatial resolution and pixel size randomness.  Therefore, we tried to remove the grid pattern by applying an fast Fourier transformation filter to the image.  The resulting image is shown in Fig.\ref{Img}(c).  Oscillations at the background are reduced.

Another standard way to remove this artifact, measurements without sample, called open-beam measurements, are necessary. 
However, open-beam measurements are hard to perform when the sample is cooled together with the detector as in the present experiments. 
Moreover, such operation produces a subtle difference between the neutron-flux distribution in open-beam imaging and that in imaging with sample. 
When the sample is placed at room temperature, open-beam measurements can be executed while the spatial resolution is limited because of the relatively long distance between the sample and detector.

\subsection{Assignment of Bragg-dip indices}

We computed the neutron transmission ratios by dividing NoE summed from the areas \#1 and \#2 of Fig.~\ref{Img}(b) by NoE summed from the area without YbSn$_3$ samples. 
We assume that the detection efficiencies and neutron flux are homogeneous over these areas to estimate the transmission ratios from the above procedure. 
We discuss resonance absorptions in the subsection~\ref{resonance}. 
Clear observation of Bragg dips indicates single-crystal growth of YbSn$_3$ owing to a high neutron scattering cross-section in natural Yb nuclei.

\begin{figure}
\begin{center}
\includegraphics[width=0.8\linewidth, pagebox=cropbox, clip]{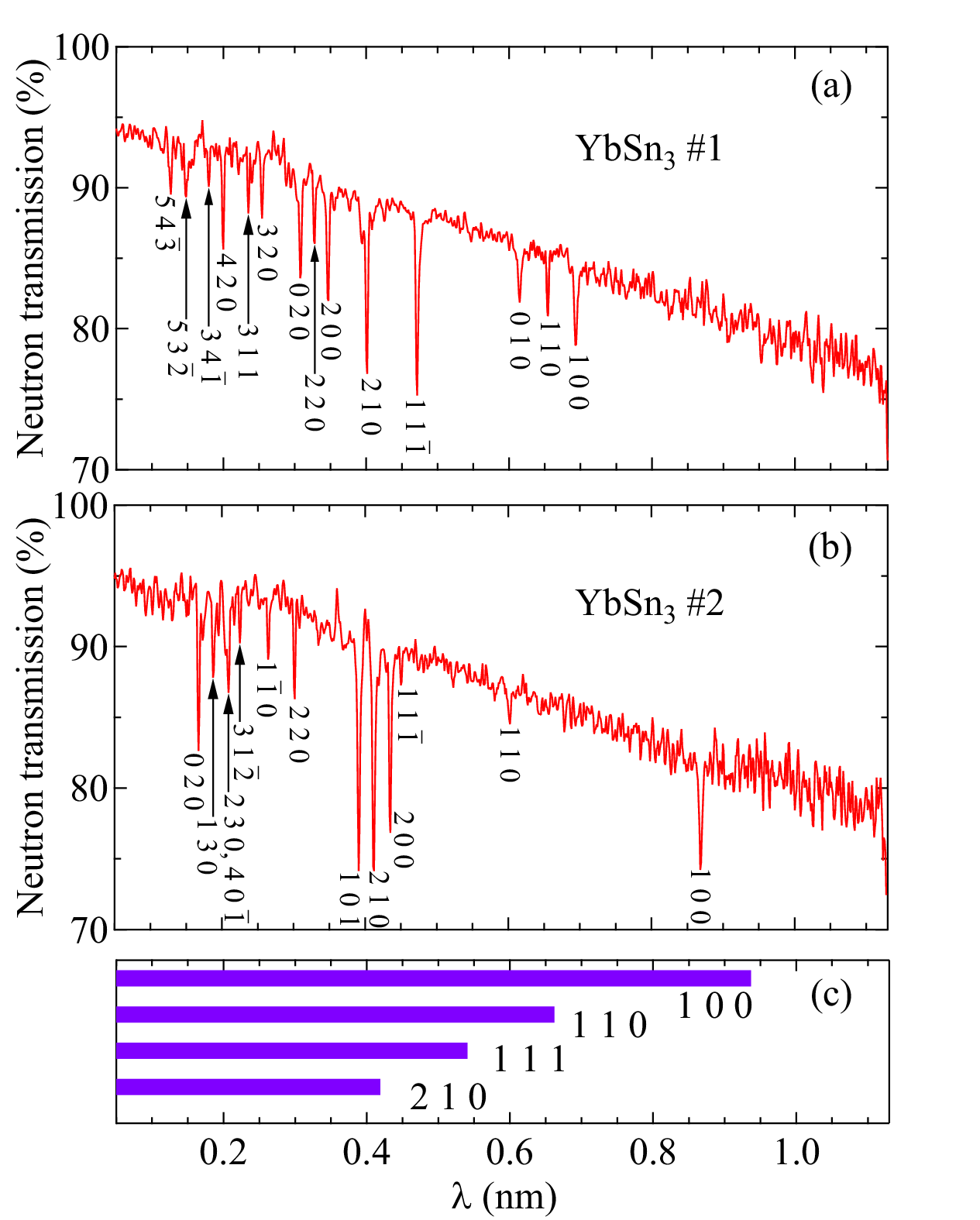}
\end{center}
\caption{(a) (b) Neutron transmissions of YbSn$_3$ summing areas \#1 and \#2 surrounded by red lines in Fig.~\ref{Img}(b). 
(c) Neutron wavelength range available to appear for each Bragg dip. }
\label{Trans}
\end{figure}

Figures~\ref{Trans}(a) and (b) show that all resulting Bragg dips of single crystals \#1 and \#2 are indexed by YbSn$_3$ diffractions. 
The difference in dip positions between \#1 and \#2 indicates that the crystal orientation of one single crystal differs from the other. 
As discussed previously \cite{Mal16}, Bragg-dip depths are related to a Laue experiment, and not all Bragg reflections are excited for a given crystal orientation.

Figure~\ref{Trans}(c) shows the available $\lambda_{hkl}$ range for each $hkl$ Bragg reflections. 
The Bragg dip at $\lambda$=0.693\,nm(0.868\,nm) in Fig.~\ref{Trans}(a)((b)) is immediately assigned as the 100 for crystal \#1(crystal \#2), because it is longer than the maximum wavelength of $\lambda_{110}$. 
In the same manner, the Bragg dip at $\lambda$=0.602\,nm in Fig.~\ref{Trans}(b) is assigned as the 110 for crystal \#2.
The 110 Bragg dip for crystal \#1 is either $\lambda$=0.654 or 0.614\,nm(see Fig.~\ref{Trans}(a)). 
We assumed the Bragg dip at $\lambda$=0.654\,nm as the 110 Bragg dip, and thus successfully assigned all Bragg dips.
The orientation of the [100] direction of the crystal \#1 was determined as $\phi$=41.0\,degrees and $\psi$=169.0\,degrees. 
That of the crystal \#2 was determined as $\phi$=20.9\,degrees and $\psi$=173.3\,degrees.
The angles $\phi$ and $\psi$ are defined in Fig.~\ref{Ewald}.

\subsection{Bragg-dip-selective neutron transmission imaging}

\begin{figure}
\begin{center}
\includegraphics[width=0.8\linewidth, pagebox=cropbox, clip]{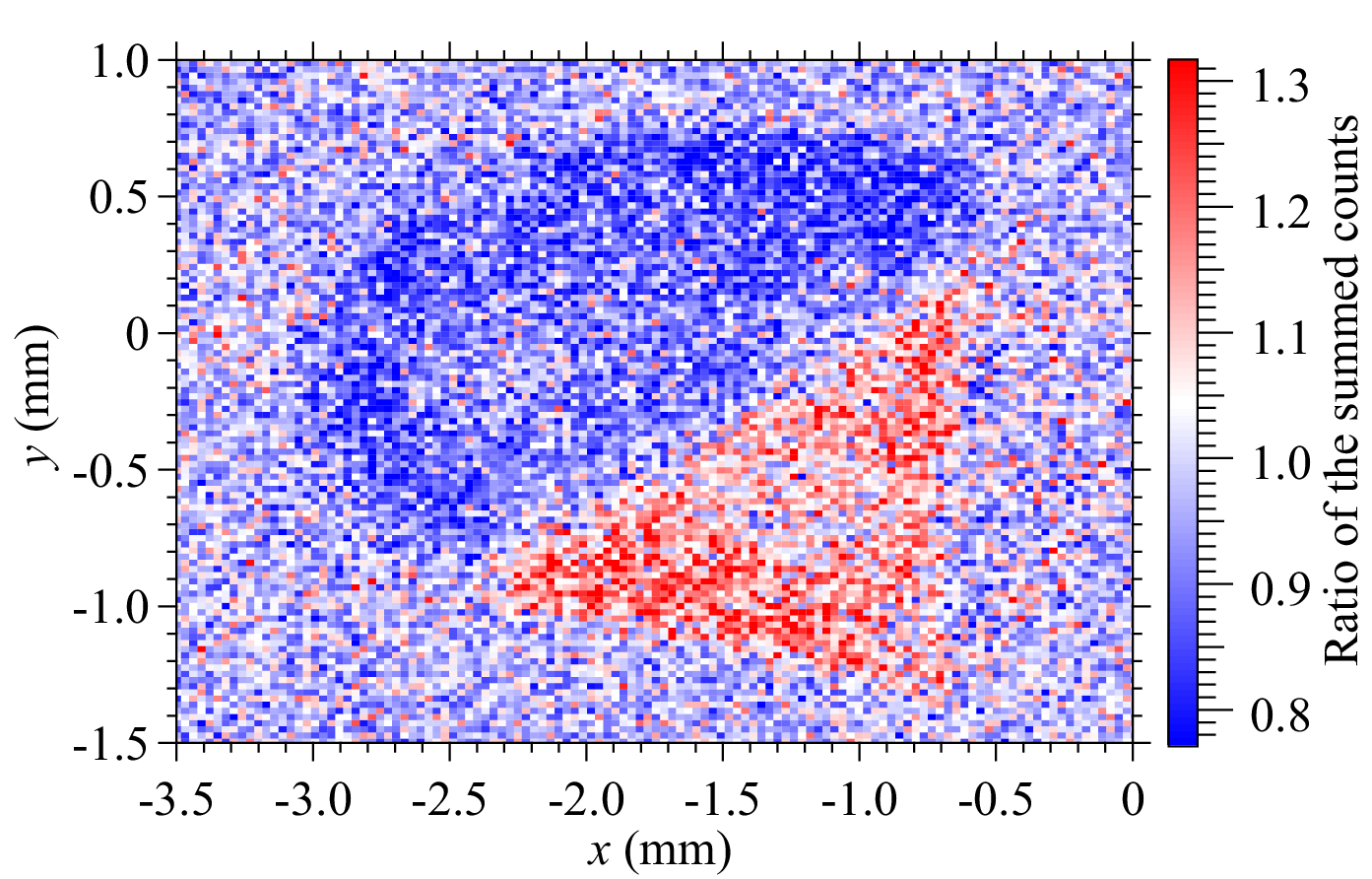}
\end{center}
\caption{Neutron transmission image emphasizing the difference in Bragg-dip positions between crystals \#1 and \#2 (see text for how to obtain the image).  }
\label{BDI}
\end{figure}

Figure~\ref{BDI} shows a neutron transmission image emphasizing the difference in Bragg-dip positions between single crystals \#1 and \#2 by choosing the pixel size as 30 $\times$ 23\,$\mu$m$^2$.  
Meanwhile, the total NoE (denoted NoE\#1) is summed up over $\lambda =$ 0.345 to 0.348\,nm (the 200 dip), 0.399 to 0.402\,nm (the 210 dip), and $\lambda=$ 0.469 to 0.474\,nm (the 11$\bar{1}$ dip) for the crystal \#1.
The total NoE (denoted as NoE\#2) is summed up over $\lambda=$ 0.389 to 0.393\,nm (the 101 dip), 0.410 to 0.414\,nm (the 210 dip), and 0.433 to 0.437\,nm (the 200 dip) for the crystal \#2. 
The ratio NoE\#1/NoE\#2 is mapped as a color image in Fig.~\ref{BDI}.
Single crystals can clearly be recognized in color as the separate blue (\#1) and red (\#2) regions.
Additionally, the image is rather uniform in the colored region of each crystal.  
Therefore, we conclude from Fig.~\ref{BDI} that the YbSn$_3$ sample consists of two single crystals with different orientations from each other.  

It is possible that several single crystals are stuck together to appear as a single crystal at first glance.
Therefore, this method is useful for characterizing the quality of single crystal samples in material sciences.
We aimed to demonstrate the usefulness of the Bragg-dip analyses in characterizing the test sample. 
We did not take into account the dip depths in this study.  
Analysis that takes into account dip depth is a subject for future work.

\subsection{Resonance absorptions of Yb nuclei\label{resonance}}

\begin{figure}
\begin{center}
\includegraphics[width=0.8\linewidth, pagebox=cropbox, clip]{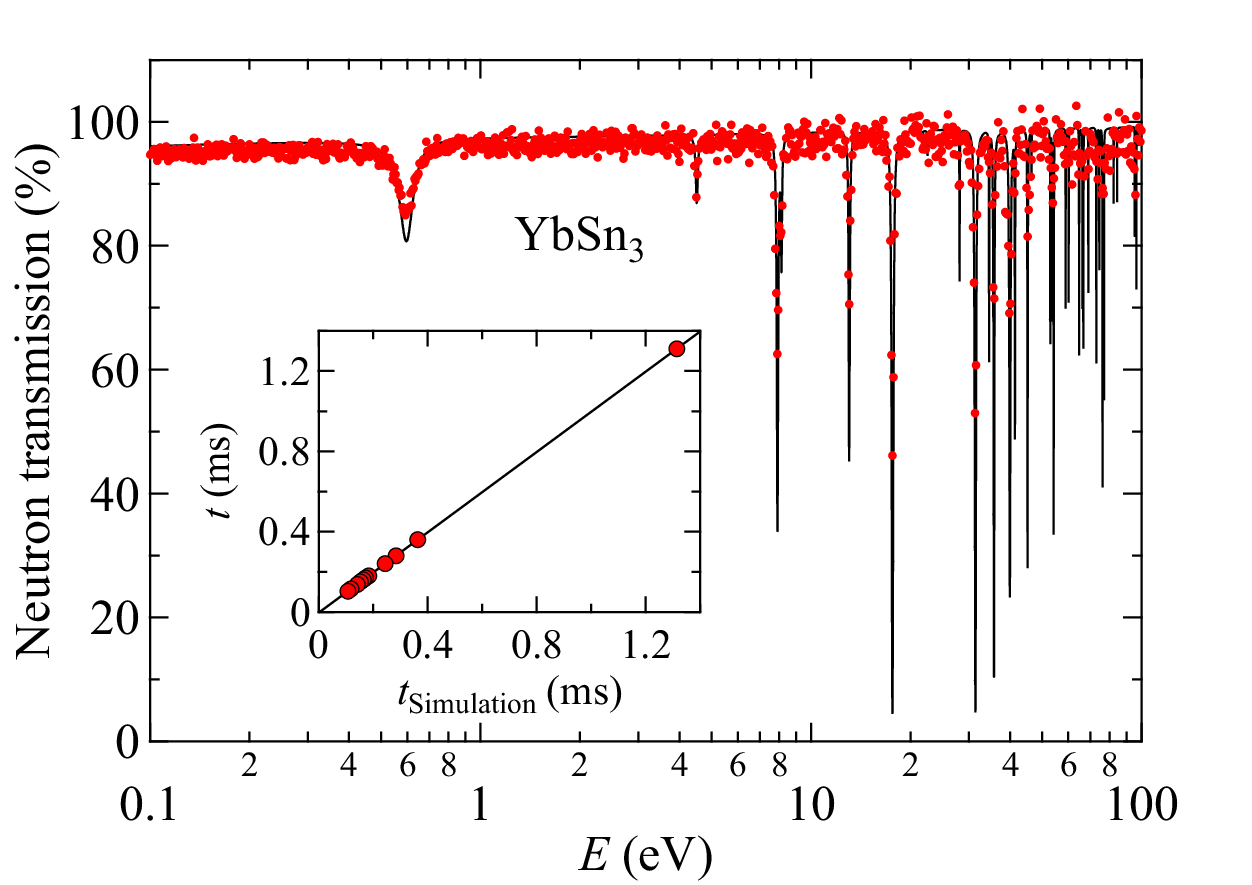}
\end{center}
\caption{Neutron transmission spectrum as a function of the neutron energy $E$. The solid line represents a simulated transmission spectrum by assuming natural Yb nuclei in YbSn$_3$. 
The inset shows the comparison between experimental and simulated dip energies.}
\label{Trans_logE}
\end{figure}

Figure~\ref{Trans_logE} shows the energy dependence of the transmission for YbSn$_3$ crystals computed by dividing NoE summing over the area \#3 of Fig.~\ref{Img}(b) by NoE summing over the area without containing samples. 
Several dips were observed by nuclear resonance absorption between Yb nuclei and neutrons. 
Here neutron energy of Bragg dips shown in Figs.~\ref{Trans}(a) and (b) are lower than the minimum range of 0.1\,eV in Fig.~\ref{Trans_logE}.
We calibrated a $t_{\rm delay}$ by comparing these dips with those in the database \cite{JENDL}. 
As shown in the inset of Fig.~\ref{Trans_logE}, the experimental ToF dips and those in the database exhibit a linear relationship with $t_{\rm delay}$=3.13\,$\mu$s and a slope of 1.00. The slope being close to 1 indicates that the actual beamline length is equal to 14\,m. 
Thus, the transmission spectrum is well reproduced by the simulation for resonance absorption by Yb nuclei as shown by the solid line in the main panel of Fig.~\ref{Trans_logE}.

\section{Summary}
In this study, we demonstrated that the delay-line CB-KID could be used for Bragg-dip and resonance-dip analyses in neutron transmission spectrum and imaging. 
Observed Bragg dips were successfully assigned to indexed Bragg reflections of two separate YbSn$_3$ single crystals. 
We determined the [100] axis orientations and created neutron images that identified the individual single crystals.   
Additionally, we also observed dips induced by resonance absorption of Yb nuclei up to 100\,eV. 
The delay-line CB-KID is a $^3$He-free neutron detector with high temporal resolution. 
The combination of the delay-line CB-KID and high-power pulsed neutron source allows us to measure the neutron-transmission spectrum and to image over a wide energy range of neutrons.

\ack{Acknowledgements}
This work is partially supported by Grant-in-Aid for Scientific Research (Nos. JP16H02450, JP21H04666, JP21K14566 and JP23K13690) from JSPS. The neutron irradiation experiments at the Materials and Life Science Experimental Facility (MLF) of the J-PARC were conducted under the support of MLF project program (No. 2020P0201).

\bibliographystyle{iucr} 
\bibliography{CBKID_YbSn3_v2}

\newpage


\end{document}